\documentclass[prc,unsortedaddress,superscriptaddress,amsmath,amsfonts,amssymb,showpacs,floatfix]{revtex4}
\usepackage{graphics,epsfig}
\begin{document}

\title{Coulomb breakup of $^6\rm{Li}$ into $\alpha\,+\,d$ in the field of a $^{208}\rm{Pb}$ ion}

\author{B.F. Irgaziev}\altaffiliation[\textit{Permanent address}:]{\,Institute of Applied Physics, NUUZ, Tashkent, Uzbekistan.}
\email{irgaziev@yahoo.com}
\author{Jameel-Un Nabi}
\author{Darwaish Khan}
\affiliation{GIK Institute of Engineering Sciences and Technology,
Topi, Pakistan}

\begin{abstract}
The triple differential cross section of the
$^{208}\rm{Pb}(^6\rm{Li},\alpha d)^{208}\rm{Pb}$ quasielastic
breakup is calculated at a collision energy of 156 MeV and a
scattering angle range of $2^\circ$-$6^\circ$. We fit the
parameters of the Woods-Saxon potential using the experimental
$\alpha$-$d$ phase shifts for different states to describe the
relative motion of $\alpha$ particle and deuteron. To check the
validity of the two particle approach for the $\alpha$-$d$ system,
we apply a potential model to describe the
$^2\rm{H}(\alpha,\gamma)^6\rm{Li}$ radiative capture. We calculate
the Coulomb breakup using the semiclassical method while an
estimation of the nuclear breakup is made on the basis of the
diffraction theory. A comparison of our calculation with the
experimental data of Kiener \textit{et al.} [ Phys. Rev. C \textbf{44}, 2195 (1991)] gives evidence for
the dominance of the Coulomb dissociation mechanism and the
contribution of nuclear distortion, but is essentially smaller
than the value reported by Hammache \textit{et al.} [ Phys. Rev. C \textbf{82}, 065803 (2010)]. The
results of our calculation for the triple differential cross sections
(contributed by the Coulomb and nuclear mechanisms) of the
$^6\rm{Li}$ breakup hint toward a forward-backward asymmetry in
the relative direction of the $\alpha$-particle and deuteron
emission, especially at smaller scattering angles, in the
$^6\rm{Li}$ center-of-mass (c.m.) system.
\end{abstract}

\maketitle

\section{Introduction}
\label{int} The study of processes, relevant for nuclear
astrophysics, by indirect methods gives the possibility to extract
the astrophysical $S$ factor at extremely low energies when
extraction by direct methods is not possible due to the Coulomb
barrier suppression. Among these indirect methods we cite the
elastic Coulomb dissociation method suggested by Baur, Bertulani
and Rebel \cite{bbr86,bert88,baur03}, the asymptotic normalization
coefficient method (ANC) suggested by Mukhamedzhanov and Timofeyuk
\cite{akram90,aktim90,xu94}, and the Trojan horse method (THM)
suggested by Baur and modified by Spitaleri \cite{baur86,spit01}.
The study of nuclear reactions at high energy is, in general, very
complicated owing to the strong nuclear interaction between the
colliding nuclei. However, in the peripheral collisions of a light
nucleus with heavy target the reaction mechanism becomes simple
owing to the negligible contribution of nuclear distortion and
excitation becomes purely Coulombic. Electromagnetic excitation is
a very powerful tool for the extraction of the information
concerning radiative capture at extremely low energies when the
direct measurements of the radiative capture of nuclei for
astrophysical purposes is impossible. The Coulomb dissociation
experiments are being performed at different centers around the
world (e.g. GSI, Germany \cite{gut01}; MSU/NSCL, USA \cite{usa00};
RIKEN, Japan \cite{riken}). The value of the astrophysical $S$
factor extracted from the Coulomb breakup experiment is affected
by various uncertainties, namely: (i) the method of extrapolation
of the data to zero energy; (ii) contributions from various
electromagnetic multipoles ($E1,\,E2,\,M1$); (iii) assumptions
about the nuclear interactions; and, (iv) various higher-order
effects (see Refs.
\cite{esbensen95,bertulani95,typel01,banerjee02} and references
therein). The astrophysical $S$ factor at zero energy extracted
from the direct radiative capture and the Coulomb breakup reaction
should be the same. However, the extrapolation to zero energy from
the $d(\alpha,\gamma)^6\rm{Li}$ and
$^{208}\rm{Pb}(^6\rm{Li},\alpha\, d) ^{208}\rm{Pb}$ reactions
gives different values for the astrophysical factor. This is owing
to the fact that at $\alpha d$ relative energy below 100 keV the
$E1$ dipole cross section becomes larger than the $E2$ quadrupole
cross section for the direct capture process, while for the
Coulomb breakup the $E1$ cross section is always smaller than the
$E2$ one at any given energy. Kiener \textit{et al.} \cite{kin91}
investigated the Coulomb breakup of $^6\rm{Li}$ in the field of
$^{208}\rm{Pb}$ ion and extracted large value of the astrophysical
factor at zero energy from extrapolation in the range $E_{\alpha
d}<400$ keV. Shyam \textit{et al.} \cite{shaym91} underlined that
the result of the authors of Ref. \cite{kin91} was free from the
nuclear background which maybe important. Kiener \textit{et al.}
did not take the contribution of the dipole transition to the
cross section owing to isospin selection rule. However, this rule
is violated and $E1$ transition can still occur. Recently a new
measurement of the Coulomb breakup at extremely high energy  of
$^6\rm{Li}$  (150 A MeV)  was performed \cite{hamach10} and the
authors have claimed disclosure of evidence for the large
contribution of the Coulomb-nuclear interference. However, the
authors  did not present the cross sections (the histograms for
counts in Figs. 6 and 7 of Ref. \cite{hamach10} were shown).
Figure 10 of Ref. \cite{hamach10} showed the ratio of nuclear to
Coulomb differential cross sections for $^6\rm{Li}$ which is very
large compared to the qualitative estimation given earlier in Ref.
\cite{bert91}.

The dissociation by the nuclear field of a target can be excluded
by observation of the fragments of the reaction at forward
scattering angles. This extremely small angle corresponds to the
impact parameter essentially larger than the sum of the nuclear
radii of the projectile and the target. For instance, in the
experiments \cite{kin91} the angle was varied in the range
$2^\circ$-$6^\circ$ corresponding to the impact parameter in the
range $20$-$65$ fm. However, the dipole transition in the
$^6\rm{Li}$ Coulomb breakup is suppressed as in the
$d(\alpha,\gamma)^6\rm{Li}$ radiative capture, therefore, we can
expect that the nuclear dissociation gives a relatively large
contribution to the $^6\rm{Li}$ Coulomb breakup cross section.

The problem of the higher-order effects was discussed in many
papers (see, e.g. Refs. \cite{baur03, bert08}, and references
therein). In the semiclassical theory these effects are inversely
proportional to the impact parameter and the velocity of
collision. Therefore we can constrain ourselves by the first-order
amplitude for the excitation of a fast nucleus at sufficiently
small scattering angles. Among higher-order effects, we also note
the three-body Coulomb effects in the final state, which were
discussed in Refs. \cite{alt03,alt05}.

In this paper we consider the dissociation
$^{208}\rm{Pb}(^6\rm{Li},\alpha\, d) ^{208}\rm{Pb}$ using the
time-dependent perturbation theory for the Coulomb breakup,
whereas the nuclear breakup is viewed as a diffractive
dissociation. Our treatment is used for the complete analysis of
the experiments, which were performed with a 156 MeV $^6\rm{Li}$
beam at the Karlsruhe Isochronous Cyclotron \cite{kin91}. This
reaction is relevant to the $d(\alpha,\gamma)^6\rm{Li}$ radiative
capture, which is one of the important nucleosynthesis reactions.
When the collision energy is sufficiently large, we can consider
the motion of the center of mass of a projectile along a classical
trajectory. At small scattering angles of the projectile the value
of the impact parameter would be so large that at such distances
only the Coulomb interaction between the projectile and the target
would be significant. The nuclear breakup does take place, but it
mainly occurs near the target surface where the nuclear potential
of the interaction falls down quickly. At small scattering angles
and high energy collisions, the most appropriate method of
estimation of the nuclear breakup is the diffraction theory
\cite{sit57} as also has been mentioned above. Here we mention the
experiments concerning the ${}^6\rm{Li}$ breakup at energies 60
MeV \cite{hess91} and 31, 33, 35 and 39 MeV \cite{maz03}, which
also confirmed the Coulomb dissociation mechanism of ${}^6\rm{Li}$
on $^{208}\rm{Pb}$. However, from our opinion the authors of Ref.
\cite{hess91} made an erroneous conclusion that the second-order
Coulomb excitation theory can improve the deviation in the angular
distribution between the theory and the experiment. On the basis
of the developed method we desire to make the analysis of the data
presented in Ref. \cite{kin91} and to explore the possibility of
the extraction of the astrophysical $S$ factor from the
$^{208}\rm{Pb}(^6\rm{Li},\alpha\, d)^{208}\rm{Pb}$ reaction by a
comparison of our results with the experimental data of Ref.
\cite{kin91}.

We use the system of units in which $\hbar=c=1$.

\section{General formalism}
\label{gen}

In the semiclassical theory the center-of-mass (c.m.) motion of
the projectile is considered classically, whereas the relative
motion of the clusters in the projectile is treated completely in
a quantum mechanical fashion \cite{baur03}. In this approach the
purely Coulomb breakup cross section of peripheral
$^{208}\rm{Pb}(^6\rm{Li},\alpha d)^{208}\rm{Pb}$ reaction is
presented as a product of the Rutherford scattering cross section
of $^6\rm{Li}$ in the field of $^{208}\rm{Pb}$ ion and the
probability of the $^6\rm{Li}\rightarrow \alpha+d$ disintegration.
We calculate the probability of disintegration in the $^6\rm{Li}$
frame (projectile frame) because the calculation does not depend
on any reference frame. In this frame a heavy $^{208}\rm{Pb}$
moves along a straight line with a constant velocity {\bf v}. If
the energy of the $\alpha d$ relative motion is small we can
restrict by using only the low partial waves at $l=0,1,2$ for the
description of the $\alpha d$ motion. The $E2$ multipole gives the
main contribution to the transition amplitude of the
electromagnetic dissociation of the $^6\rm{Li}$, however, the $E1$
transition should also be included to the amplitude owing to a
violation of the isospin forbidden rule.

The time-dependent perturbation for the Coulomb breakup $A+a
\rightarrow  A+c+b$ is
\begin{equation}
H(t)=\int d^3{x}\Bigl(\frac{Z_A e}{\mid{\bf{x}}-{\bf{R}}(t)\mid}-
\frac{Z_Ae}
{\mid{\bf{R}}(t)\mid}\Bigr)\rho({\bf{x}}),\label{pert1}
\end{equation}
where $Z_A e$  is the charge of a heavy ion $A$ ($^{208}\rm{Pb}$
target), the projectile $a$ ($^6\rm{Li}$ nucleus) is dissociated
into $b$ (deuteron) and $c$ ($\alpha$ particle), ${\bf{R}}(t)={\bf
b}+{\bf v}t$ gives the position of the target in the projectile
frame, and $\rho({\bf{x}})$ is the charge density operator, {\bf
b} is the impact parameter. For the peripheral reaction we can
take the charge density operator in the two-body approach as
\begin{equation}\label{rho}
\rho({\bf x})=Z_be\delta({\bf x}-{\bf r}_b)+Z_ce\delta({\bf
x}-{\bf r}_c),
\end{equation}
where ${\bf r}_i$ defines the position of particle $i$ in the
projectile frame, and $Z_ie$ is its charge.

In perturbation theory the amplitude of the transition from the
initial state $|i>$ (wave function of $a=b+c$ system in the ground
state) to the final state $|f>$ (wave function of the $b+c$ system
in the continuum state) is given as a sum
\begin{equation}\label{pert2}
a_{fi}=\delta_{fi}+a_{fi}^{(1)}+a_{fi}^{(2)}+\cdots ,
\end{equation}
of different order contributions. We use the first-order constrain
for the amplitude of the transition. Expanding $H(t)$ in
multipoles we get
\begin{equation}\label{pert5}
a_{fi}^{(1)}=\frac{4\pi Z_A e}{i}\sum\limits_{\lambda,\mu}\frac{(-1)^{\mu}}{2\lambda+1}\langle f\mid{\cal{M}}(\lambda,-\mu)\mid i\rangle S_{\lambda \mu}(\omega),
\end{equation}
where $\cal{M}(\lambda,\mu)$ is the electric multipole operator
\begin{equation}\label{pert6}
{\cal{M}}(\lambda,\mu)=\int d^3x\rho({\rm{\bf x}})x^\lambda
Y_{\lambda \mu}(\hat{\bf x}).
\end{equation}
The semiclassical orbital $S_{\lambda\mu}(\omega)$ integral is
given by
\begin{equation}\label{pert7}
S_{\lambda\mu}(\omega)=\int\limits_{-\infty}^{+\infty}dt\frac{e^{i\omega
t}}{R(t)^{\lambda+1}}Y_{\lambda\mu}\bigl[\hat{\bf R}(t)\bigr],
\end{equation}
where $\hat{\bf x}$ and $\hat{\bf R}(t)$ are the unit vectors
along the position vectors ${\bf{x}}$ and ${\bf{R}}(t)$,
respectively. The integrals $S_{\lambda\mu}(\omega)$ were
calculated analytically  and the results of calculations were
presented in Ref. \cite{esb02}. Determining the relative
coordinate ${\bf r}={\bf r}_b-{\bf r}_c$, inserting Eq.(\ref{rho})
into Eq.(\ref{pert6}) and performing the integration over variable
${\bf x}$ we obtain
\begin{equation}\label{pole}
{\cal{M}}(\lambda,\mu)=\mu_{bc}^\lambda\Bigl[\frac{Z_be}{m_b^\lambda}+(-1)^\lambda\frac{Z_ce}{m_c^\lambda}\Bigr]r^\lambda
Y_{\lambda \mu}(\hat{\bf r}),
\end{equation}
where $\mu_{bc}$ is the reduced mass of particles $b$ and $c$. We
see that the dipole transition operator ${\cal{M}}(1,\mu)$ is not
equal to zero because the value of the charge-to-mass ratio is
slightly different for the $\alpha$ particle and deuteron.
Therefore, the $E1$ transition amplitude is not zero.

The triple cross section of the Coulomb breakup from the  ground
state with angular momentum and parity $J_i^\pi=1^+$ of
$^6\rm{Li}$ to the final state with relative momentum $\bf{k}$ of
the $\alpha$ particle and deuteron having reduced mass
$\mu_{\alpha d}$ can be expressed in terms of the excitation
amplitude $a_{fi}$:
\begin{equation}\label{pert8}
\frac{d^3\sigma_C}{d\Omega_{\alpha d}d\Omega_{\rm{Li}}dE_{\alpha
d}}=\frac{d\sigma_R}{d\Omega_{\rm{Li}}}\frac{1}{2
J_i+1}\sum\limits_{M_i}\mid a_{fi}\mid^2\frac{\mu_{\alpha
d}k}{(2\pi)^3}.
\end{equation}
The elastic Coulomb cross section $d\sigma_R/d\Omega_{\rm{Li}}$ is
calculated classically for the scattering of the c.m. of the
projectile $^6\rm{Li}$.

Next we consider the nuclear breakup. As mentioned earlier, the
nuclear breakup occurs mostly near the surface of a heavy target
nucleus $^{208}\rm{Pb}$. If the scattering angle is small we may
apply the diffraction theory assuming that the target nucleus is a
completely absorptive ``black'' sphere of radius $R_{bl}$. The
amplitude for the elastic breakup according to the diffraction
theory \cite{rebel90} is given by
\begin{equation}\label{nucl-ampl}
{\cal F}({\bf q, k})=\frac{i k_i}{2\pi}\int d^2 be^{i{\bf
q\,b}}\int d^3 r\psi^{*}_{\bf k}({\bf r})\omega({\bf b, r
})\psi_0({\bf r}),
\end{equation}
where ${\bf q}={\bf k}_i-{\bf k}_f$, ${\bf k}_i$ is the initial
momentum of $^6{\rm Li}$ and ${\bf k}_f$ is the final momentum of
the center of mass  of the ${\alpha\, d}$ system after
dissociation; $\mid i\rangle=\psi_0({\bf r})$ and $\mid
f\rangle=\psi_{\bf k}({\bf r})$ are the wave functions of
${\alpha\, d}$ in the bound and continuum states, respectively;
{\bf b} is the impact parameter of the ${\alpha\, d}$ system;
$\omega({\bf b, r })$ is the total profile function for the
${\alpha\, d}$ system. We may take the vector ${\bf q}$ to be
orthogonal to the momentum vector ${\bf k}_i$ due to high energy
collision and small scattering angle ($q\approx k_i\cdot\theta$,
where $\theta$ is the scattering angle). The total profile
function
\begin{equation}\label{omega}
\omega=\omega_{\alpha}+\omega_d-\omega_{\alpha}\omega_d
\end{equation}
is composed of the profile functions of the fragments. The third
term in Eq. (\ref{omega}) describes the double scattering and its
contribution to the cross section is much smaller than the
contributions of the first two terms. Neglecting the
double-scattering term we obtain the cross section of the nuclear
breakup as
\begin{equation}\label{cross-nucl}
\frac{d^3\sigma_{N}}{d\Omega_{\alpha d}d\Omega_{\rm{Li}}dE_{\alpha
d}}=\mid k_i R_{\rm{bl}}\frac{J_1(q R_{\rm{bl}})}{q}s({\bf q,
k})\mid^2\frac{\mu_{\alpha d}k}{(2\pi)^3},
\end{equation}
where $J_1(x)$ is the Bessel function and
\begin{equation}\label{sqk}
s({\bf q, k})=\int d^3r\, e^{i{\bf q\, r}}\psi^{*}_{\bf k}({\bf
r})\psi_0({\bf r}).
\end{equation}
The total cross section  is considered as the sum
\begin{equation}\label{total}
\frac{d^3\sigma_{t}}{d\Omega_{\alpha d}d\Omega_{\rm{Li}}dE_{\alpha
d}}=\frac{d^3\sigma_{C}}{d\Omega_{\alpha
d}d\Omega_{\rm{Li}}dE_{\alpha
d}}+\frac{d^3\sigma_{N}}{d\Omega_{\alpha
d}d\Omega_{\rm{Li}}dE_{\alpha d}}.
\end{equation}
In Eq. (\ref{total}) the Coulomb-nuclear interference term is
neglected. We note that the Coulomb and nuclear breakup are
calculated using different approaches therefore we can not
calculate the interference term. Even if both the Coulomb and
nuclear breakup amplitudes were calculated using the same
approach, one expects the interference term to be small (see for
example the calculation in the distorted wave Born approximation
(DWBA) method used by Bertulani and Hussein, Figs. 4 and 5 in Ref.
\cite{bert91}). We further note that at the small scattering
angles considered in this calculation, the elastic Coulomb
scattering cross section calculated classically is equal to the
cross section calculated by the diffraction theory.

\section{$\alpha$-$d$ potentials}
\label{pot} To describe the relative motion of the $\alpha$
particle and deuteron we use the Woods-Saxon potential with the
orbital terms,
\begin{equation}\label{pert9}
V_N(r)=-\Bigl[V_0-V_{sl}{\bf{(l\cdot s)}}\frac{1}{m_\pi^2\,
r}\frac{d}{dr}\Bigr]\frac{1}{1+\exp\left[(r-R_N)/a\right]},
\end{equation}
with the standard value of the diffuseness $a=0.65$ fm. The
parameters $V_0$ and $V_{sl}$ for different $\alpha d$ states  is
fitted from the shift phase analysis. We take the nuclear radius
of the potential $R_N$ as
\begin{equation}\label{pert10}
R_N=r_0\cdot A^{1/3},
\end{equation}
with the standard value $r_0=1.25$ fm, and the nuclear mass number
$A=6$ for $^6\rm{Li}$ . The Coulomb potential is taken as
\begin{equation}\label{pert11}
V_{C} (r)=\left\{\begin{array}{c} {\frac{Z_{\alpha} Z_{d} e^{2} }
{2R_{C} } (3-\frac{r^{2} }{R_{C}^{2} } ),{ \; \; \; \; \; \; \; }}
r < R_{C}, \\ {\frac{Z_{1} Z_{2} e^{2} }{r}, { \; \; \; \; \; \;
\; \; \; \; \; \; \; \; \; \; \; \; \; \; }}
r>R_{C}, \end{array}\right.
\end{equation}
where $Z_{\alpha}e$ and $Z_de$ are the charges of the $\alpha$
particle and deuteron, respectively; $R_C=r_C\,A^{1/3}$
($r_C=1.25$ fm).

The parameters of the depth of the potentials $V_0$ and $V_{sl}$
were fixed by fitting the experimental \textit{S}, \textit{P} and
\textit{D} phase shifts of the elastic $\alpha$-$d$ scattering
\cite{mci67,schm72,grub75,jen83,kuk91} and the binding energy of
the $^6\rm{Li}$ ground state. We obtained the following values of
the depths: $V_0=60.73$ MeV for the $^3S_1$ state; $V_0=57.0$ MeV,
$V_{sl}=4.0$ MeV for the $^0P_1$, $^1P_1$ and $^2P_1$ states;
$V_0=55.9$ MeV, $V_{sl}=4.0$ MeV for the $^1D_1$, $^2D_1$ states
and $V_0=55.9$ MeV, $V_{sl}=5.06$ MeV for the $^3D_1$ state. To
describe the $3^+$ resonance  of  $^6\rm{Li}$ correctly, the depth
of the spin-orbital  part of the potential for $^3D_1$ state is
taken slightly differently. We note that our fitted parameters of
the Woods-Saxon potential are slightly different from the
parameters used in Ref. \cite{hamach10}. This difference is
obvious as the phase shifts are determined with errors and the
resulting fit can give rise to such differences in the parameters
of the potential. Note the spin-orbital potential in
Eq.(\ref{pert9}) contains the dimensional parameter
$1/m_\pi^2=2.136$ fm$^2$ while the same potential in Ref.
\cite{hamach10} has $\lambda^2=4$ fm$^2$.

It is important to remember that at low energies the $\alpha d$
radiative capture depends on the tail of the $^6\rm{Li}$ bound
state wave function projected  on the $\alpha$-$d$ channel
\cite{akr11,akram95}. The amplitude of this tail is the asymptotic
normalization coefficient (ANC). We note that the range for the
ANC obtained by various techniques is wide
($C=1.51$-$3.25\,\,\rm{fm}^{-1/2}$) \cite{blok93}. We calculate
the amplitude of the peripheral radiative capture reaction
$^2\rm{H}(\alpha, \gamma)^6\rm{Li}$ at very low collision energies
where the ANC for the virtual decay $^6\rm{Li} \to \alpha + d$
from the ground state governs the overall normalization of the
peripheral reaction cross section. The value of the ANC ($C_0$)
obtained here using the fitted potential for the $^3S_1$ state of
$^6\rm{Li}$ is 2.7 $\rm{fm}^{-1/2}$. This value is larger than
$C=2.3\,\, \rm{fm}^{-1/2}$ extracted from the elastic $\alpha$-$d$
$^3S_1$ experimental phase shift by the analytic extrapolation to
the pole of the partial scattering amplitude corresponding to the
$^6\rm{Li}$ ground state \cite{blok93}. The same value of the ANC
was also obtained from the solution of the three body
$\alpha$-$p$-$n$ equation \cite{kuk84}, which was used in Refs.
\cite{akr11,akram95} and confirmed recently by \textit{ab initio}
calculations \cite{navr11}. To obtain the ANC value of 2.3
$\rm{fm}^{-1/2}$ we can find the phase-equivalent potential and
the corresponding wave function by the method described in Ref.
Ref. \cite{newton} and discussed in Ref. \cite{akr11}. The
phase-equivalent potential does not change the scattering phase
shift and the binding energy, but allows one to get the needed
value of ANC. The phase-equivalent potential has the form
\begin{equation}
V_{\rm{eff}}(r)=V_N(r)-2\frac{d^2}{dr^2}f_{\rm{eff}}(r),
\end{equation}
where $f_{\rm{eff}}(r)$ means
\begin{equation}
f_{\rm{eff}}(r)=\ln\left[1+ (\lambda -1)\left(1-\int\limits_0^r
u^2(r)dr\right)\right].
\end{equation}
The corresponding new wave function of the bound state is equal to
\begin{equation}\label{eff-u}
u_{\rm{eff}}(r)=\lambda^{1/2}\frac{u(r)}{1+(\lambda -1)\int _0^r
u^2(r)dr},
\end{equation}
where $u(r)$ is the wave function obtained from the solution of
the Schr\"odinger equation with the parameters of the Woods-Saxon
potential determined from the phase-shift analysis. We note that,
to decrease the value of the ANC we have to take $\lambda>1$. For
our phase-equivalent potential and the wave function corresponding
to $C=2.3 $ fm$^{-1/2}$ the value of $\lambda$ is 1.38. [This
follows from Eq. (\ref{eff-u}): $C=\lambda^{-1/2}C_0$]. A detailed
description of this method and its application in case of the
$\alpha$-$d$ radiative capture can be found in Ref. \cite{akr11}.
Thus for calculation of the cross sections of the
${}^2\rm{H}(\alpha, \gamma){}^6\rm{Li}$ radiative capture and the
$^{208}\rm{Pb}(^6\rm{Li},\alpha\, d)^{208}\rm{Pb}$ breakup we use
the $u_{\rm{eff}}(r)$ bound wave function having $C=2.3$
fm$^{-1/2}$.

\section{Radiative capture reaction ${}^2\rm{H}(\alpha,
\gamma){}^6\rm{Li}$}\label{capture}

Experimental measurements of the cross section of the
$^2\rm{H}(\alpha, \gamma)^6\rm{Li}$ reaction at extremely low
energies are very difficult to carry out because the cross section
is the order of a few nanobarns and decreases exponentially if
energy goes to zero. The experimental results for the cross
sections of the direct $d+\alpha$ capture were measured by
Robertson \textit{et al.} \cite{rob81}  at c.m. energy range 1-3.5
MeV, by Mohr \textit{et al.} \cite{mohr94} at the resonance point
of 711 keV, and Cecil \textit{et al.} \cite{cec96} at an $\alpha
d$ c.m. energy of 53 keV. Furthermore, we mention Refs.
\cite{noll97,noll01} where the analysis of the experimental
results and theoretical calculations are given. Nollett \textit{et
al.} \cite{noll01} used a six-body approach for the calculation of
the $\alpha d$ capture. From the analysis of the results at the
$E_{\alpha d}$ energy close to zero we see  an essential
difference in the value of the astrophysical $S$ factor depending
on the applied value of the ANC. At extremely low energies the
initial wave function of the $\alpha d$ system ceases to depend on
the parameters of the nuclear potential, which is used to
calculate this wave function, since the nuclear scattering phase
shifts tend to zero. Therefore, we can replace the wave function
of the initial state of the purely Coulomb wave if $E_{\alpha
d}<100$ keV. For any radiative capture at low energy, the main
contribution comes from $E1$, $E2$ and $M1$ transitions. However,
for the $^2\rm{H}(\alpha, \gamma)^6\rm{Li}$ reaction the main
contribution comes from the $E2$ quadrupole transition at the
energy larger than 200 keV. The $E1$ transition begins to dominate
if the $E_{\alpha d}$ energy becomes less than 100 keV. The $M1$
capture remains negligible for all astrophysical interesting range
of energies.

\begin{figure}[thb]
\begin{center}
\parbox{6.7cm}{\includegraphics[width=6.7cm]{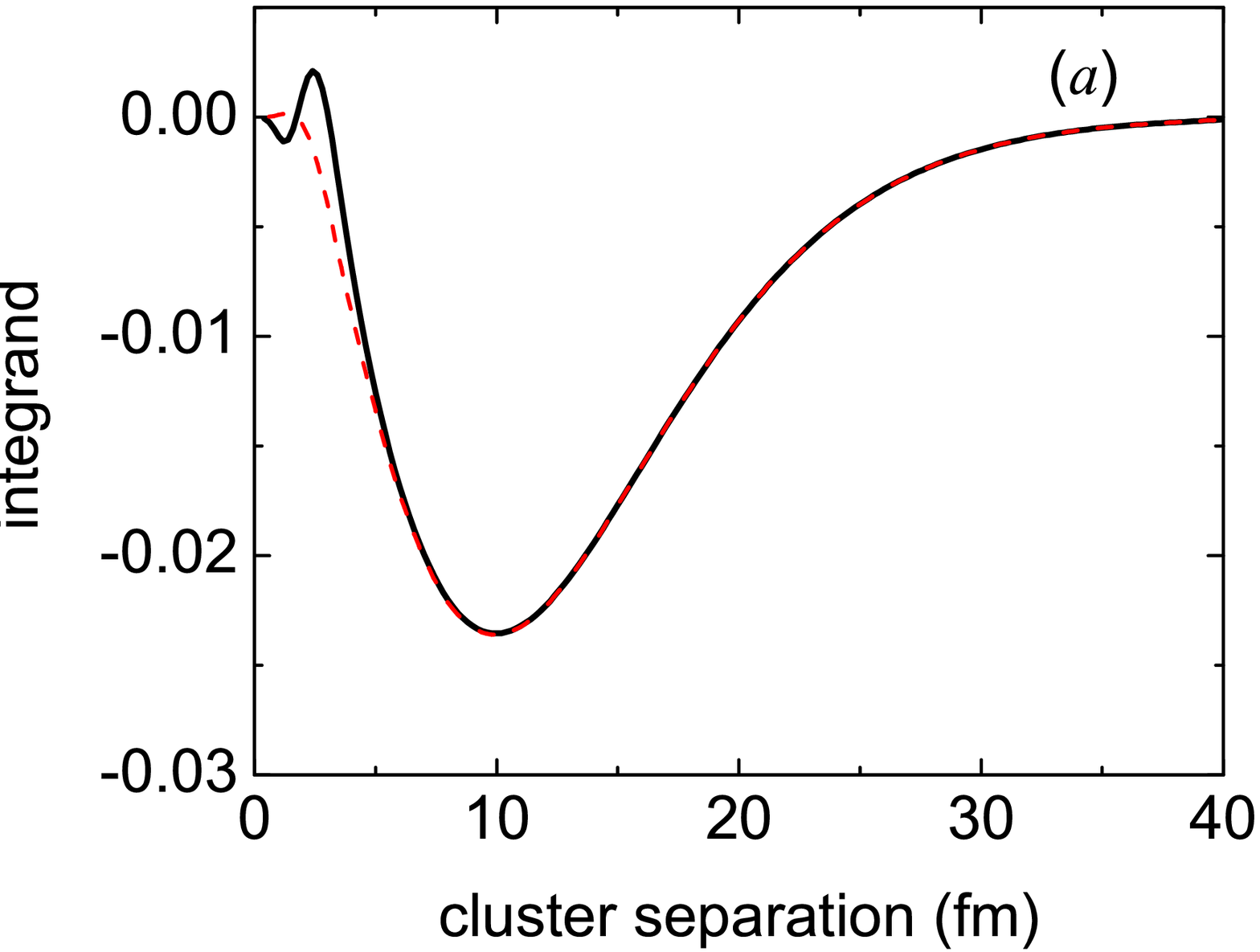}}
\parbox{6.7cm}{\includegraphics[width=6.7cm]{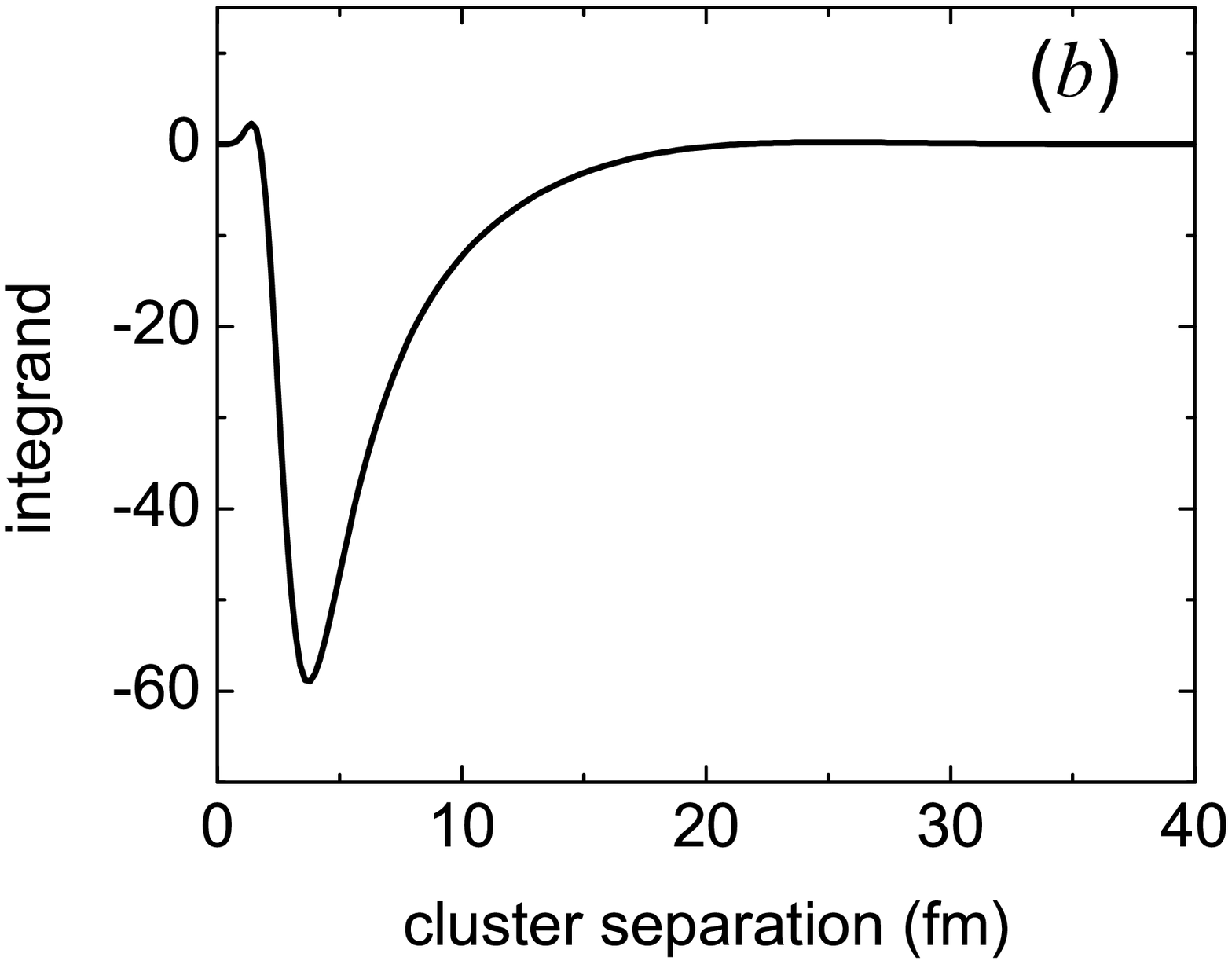}}
\end{center}
\caption{(Color online) (\textit{a}) The integrand of the matrix
element for transition from $^0P_1$ to $^3S_1$ at energy
$E_{\alpha d}=0.1$ MeV calculated with the continuum wave function
which is the solution of the Schr\"odinger equation (solid curve)
and with the regular Coulomb wave function (dashed curve).
(\textit{b}) Same as in (\textit{a}) for the transition from
$^3D_1$ to $^3S_1$ at the resonance energy $E_{\alpha d}=0.711$
MeV. \label{fig1}}
\end{figure}

For the calculation of the $d(\alpha,\gamma)^6\rm{Li}$ cross
section we use the Woods-Saxon potential with the parameters
described in the previous section. The initial wave function of
the $\alpha d$ system includes the $P$ and $D$ waves which are
solutions of the Schr\"odinger equation.  It is clear that the
$\alpha d$ direct capture is a peripheral reaction at the low
energy ($E_{\alpha d}<300$ keV). Accordingly, the cross section is
not sensitive to the choice of the parameters of potential
describing the continuum states. It is rather strongly dependent
on the ANC of the ground state. Such a property was used for the
ANC calculation \cite{akram95,akr11}, where the asymptotic wave
function of the $^6\rm{Li}$ ground state wave function in two-body
approximation was applied to find the astrophysical $S$ factor.
The main contribution to the matrix elements of the direct
reaction at $E_{\alpha d}<300$ keV comes from the external part of
the used wave functions, while the internal part gives a very
small contribution. Figure \ref{fig1}(\textit{a}) shows the radial
part of the integrand for transition from $^0P_1$ to $^3S_1$ at
the energy $E_{\alpha d}=0.1$ MeV. We can see that the replacement
of the $P$ continuum wave function by the corresponding regular
Coulomb wave function leads to almost  the same value of the
matrix element for the $^0P_1\to ^3S_1$ transition. We further
note that the same conclusion can be made for the transition from
$D$ states at the low energy.

\begin{figure}[thb]
\begin{center}
\includegraphics[width=8.6cm]{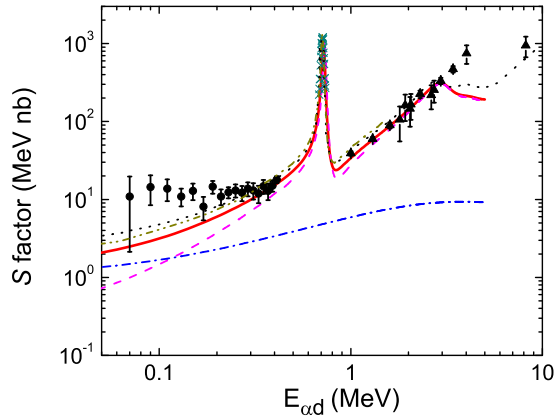}
\end{center}
\caption{(Color online) The calculated astrophysical $S$ factor
compared with the results of other theoretical calculations and
experimental data. The solid line is our result for the total $S$
factor, the dashed and dot-dashed lines are the contribution of
$E2$ and $E1$ multipolarities, respectively. The dotted line shows
the result of the calculation of the authors Ref. \cite{noll01}
and dot-dot-dashed line corresponds to Ref. \cite{hamach10}. The
experimental data ($\blacktriangle,\,\bullet,\,\times$) were taken
from their graphical presentation in Ref. \cite{noll01}.
\label{fig2}}
\end{figure}

It seems that in the resonance region the complete microscopic
model should be used (see Ref. \cite{noll01} and references
therein) because the contribution to the  transition amplitude
from the internal part of the radial wave functions is also
expected. Nevertheless, the two-particle approach can be used to
describe the $^2\rm{H}(\alpha,\gamma)^6\rm{Li}$ resonance reaction
and to get a quantitative result for the cross section at the
resonance energy region. Such instances maybe explained by the
suppression of the $E1$ transition. The resonant amplitude is the
result of the transition from the state $^3D_1$ to the state
$^3S_1$ and the transition remains peripheral due to the large
centrifugal barrier in the $^3D_1$ state.  The result of our
calculations shows that the two-body wave function for the ground
state of $^6\rm{Li}$ gives an acceptable result for the resonance
cross section and the internal part of the wave function gives
very small contribution to the transition amplitude at the
resonance energy [see Fig. \ref{fig1} (\textit{b})]. The result of
the calculations of the astrophysical $S$ factor as a function of
energy is shown in Fig.~\ref{fig2} over a wider energy range than
shown in Fig. 3 of Ref. \cite{akr11}. It is clear from
Fig.~\ref{fig2}  that the two-body approach for the description of
the ground and continuum states of the $\alpha d$ system gives
perfectly good results at the low energy, including the $3^+$
resonance energy region.
The value of our calculated astrophysical $S$ factor coincides
with the one presented in Ref. \cite{akram95} for the low
energies. At an energy larger than 1 MeV our result, definitely,
agrees well with the experimental data \cite{rob81}, while the
results of the authors of Refs. \cite{hamach10,noll01} clearly
overestimated the data.  We also see a big disagreement of our
result with the astrophysical factor extracted from the Coulomb
breakup experiment \cite{kin91} below 500 keV. We note that the
ANC determines the value of the astrophysical $S$ factor at the
energy range $E_{\alpha d}<300$ keV where the reaction has a clear
peripheral mechanism. The ratio of the astrophysical $S$ factors
(or cross sections) calculated near zero $E_{\alpha d}$ energy
using the ground wave functions with a different value of the ANC
is equal to the square of the ratio of the corresponding ANCs. As
was also mentioned above, the ground state wave function of
$^6\rm{Li}$ used in our calculation for the radiative capture has
an ANC equal to 2.3 fm$^{-1/2}$ while the ANC of the bound state
wave function used in Refs. \cite{hamach10,noll01} is 2.7
fm$^{-1/2}$ and 3.2 fm$^{-1/2}$, respectively \footnote[1]{The
authors of Ref. \cite{noll01} quoted that the two-cluster $\alpha
d$ distribution function had the dimensionless ANC $C_0=2.26\pm
0.05$ which gave the dimension ANC equaled to $1.77\pm 0.04$
fm$^{-1/2}$. This ANC must lead to $S(0)=0.72\, \rm{MeV\, nb}$ but
from Fig. 8 of Ref. \cite{noll01} we see that $S(0)\approx 2.5\,
\rm{MeV\, nb}$. Therefore we re-estimated the value of the ANC.}.
The secondary peak of the calculated cross section near $E_{\alpha
d}=3$ MeV appears due to the wide resonance in the $^3D_2$
scattering wave at $E_{\alpha d}=2.838$ MeV. The applied potential
describes this resonance. To explain the disagreement of our
results with the experimental data for $E_{\alpha d}>3$ MeV, we
refer to the work of Nollett \textit{et al.} \cite{noll01}, where
the nature of this disagreement is discussed in detail.
Additionally, we note that for the energy $E_{\alpha d}>3$ MeV the
reaction is no longer peripheral. We have not calculated the cross
section above 6 MeV because the phase shifts have a large
imaginary part due to the open $\alpha+p+n$, $^5\rm{He}+p$ and
$^5\rm{Li}+n$ channels from the energy $E > 5$ MeV and the
restriction by the single $\alpha d$ channel becomes incorrect.
From Fig. \ref{fig2} we see that the $E1$ cross section dominates
the $E2$ cross section at energies below $ 100$ keV and
$S_{E1}(0)=1.01\,\, \rm{MeV\, nb}$ while
$S_{E2}(0)=0.21\,\,\rm{MeV\, nb}$. Hence our total calculated
value of the astrophysical factor at zero energy is
$S(0)=1.22\,\,\rm{MeV\,nb}$.

\section{Results for the Coulomb breakup of $^6\rm{Li}$}
 \label{breakup}

Using the $E1$ and $E2$ multipole matrix elements calculated to
determine the $\alpha d$ capture cross section, we compute  the
$^{208}\rm{Pb}(^6\rm{Li},\alpha\, d)^{208}\rm{Pb}$ breakup
reaction. The calculated triple differential cross section
[Eq.(\ref{pert8})] includes $E1$, $E2$ and $E1E2$ terms. The
analyzed experimental data of the Coulomb breakup were taken from
Ref. \cite{kin91}, where data are presented for the scattering
angles $\Theta_{\rm{lab}}=2, 3, 4$, and  $6^\circ$ in the
laboratory frame in Tables III theough VI; and Figs.7 and 10,
respectively. To convert the laboratory cross sections into the
$\alpha d$ c.m. cross sections we used transformation method
described in Ref. \cite{fuc82}.  Previously, only the data at the
scattering angle $\Theta_{\rm{lab}}=3^\circ$ were analyzed and the
astrophysical factor at zero energy was extracted from that
analysis \cite{kin91}. The experimental data at other scattering
angles have not been analyzed so far to the best of our knowledge.
We chose the impact parameter corresponding to the selected
scattering angles when we calculate the Coulomb breakup
contribution. In spite of the fact that the $E1$ cross section
becomes larger than the $E2$ cross section for the direct
radiative capture at energies $E_{\alpha d}<100$ keV, the $E1$
triple cross section is always less than the $E2$ one for the
Coulomb breakup at any energy including the range $E_{\alpha
d}<100$ keV. In the Coulomb breakup, like the radiative capture
reaction, the contribution of $E1$ transition cross section
decreases more slowly than $E2$ cross section with reduction of
energy. For instance, at an $\alpha d$ energy of 100 keV  the the
$E1$ contribution is less than the $E2$ one by a factor 70, and at
$E_{\alpha d}=10$ keV their ratio is equal to 1/24. This means
that the Coulomb breakup cross section is mostly defined by the
$E2$ quadrupole transition at the extremely low energy as well.
There is a contribution of the $E1E2$ interference term, but the
contribution of the $E1$ term to the cross section is very small
as compared to the $E1E2$ interference term.

When we calculate the nuclear breakup cross section according to
Eq.(\ref{cross-nucl}), the radius $R_{\rm{bl}}$ of the ``black''
nuclei $^{208}\rm{Pb}$ is fixed by means of $\chi^2$ so, that the
sum of the Coulomb and nuclear cross sections is close to the
experimental data in the energy region $0.4$ MeV $<E_{\alpha
d}<0.73$ MeV. For the calculation of $\chi^2$ we take into account
the region where deutron emission occurs in the forward direction
in the $^6\rm{Li}$ c.m. frame. We do not include the region,
$E_{\alpha d}<0.4$ MeV, because in this region higher-order
effects, such as the three-body Coulomb effects
\cite{alt03,alt05}, can give a significant contribution. In the
region $E_{\alpha d}>0.73$ MeV the behavior of the experimental
cross section is not clear for the experimental scattering angles
and the experimental error is comparable with the value of the
cross section (see Tables III through VI of Ref. \cite{kin91}).
The value of $\chi^2$ has a minimum when $R_{bl}$ is taken equal
to 6.30, 6.94, 6.34, and 6.25 fm corresponding to angles
$\Theta_{\rm{lab}}=2^\circ,\,3^\circ,\,4^\circ,$ and $6^\circ$,
respectively. We see that the values of $R_{bl}$ are close to the
radius of $^{208}\rm{Pb}$.

\begin{figure}[thb]
\begin{center}
\parbox{6.7cm}{\includegraphics[width=6.7cm]{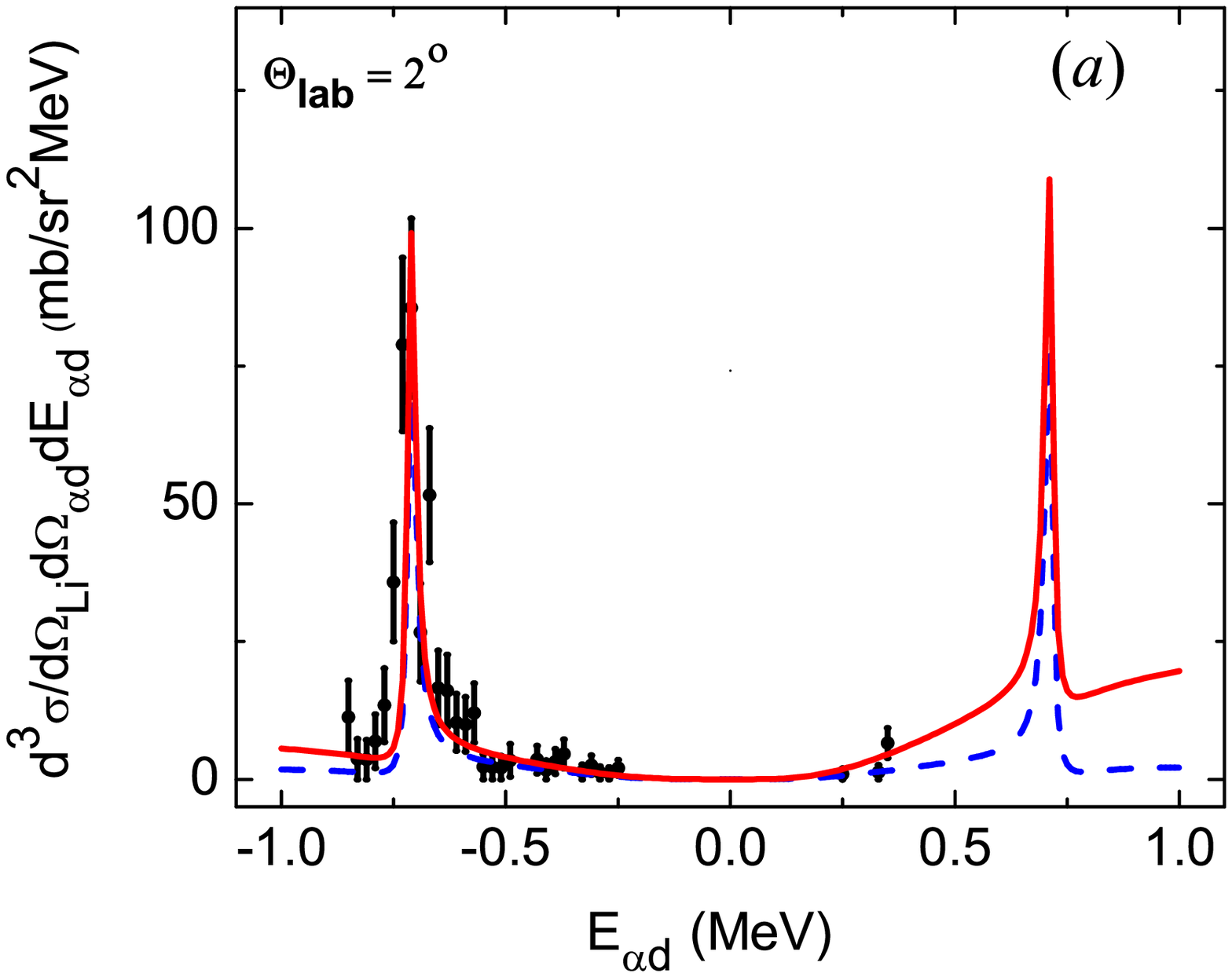}}
\parbox{6.7cm}{\includegraphics[width=6.7cm]{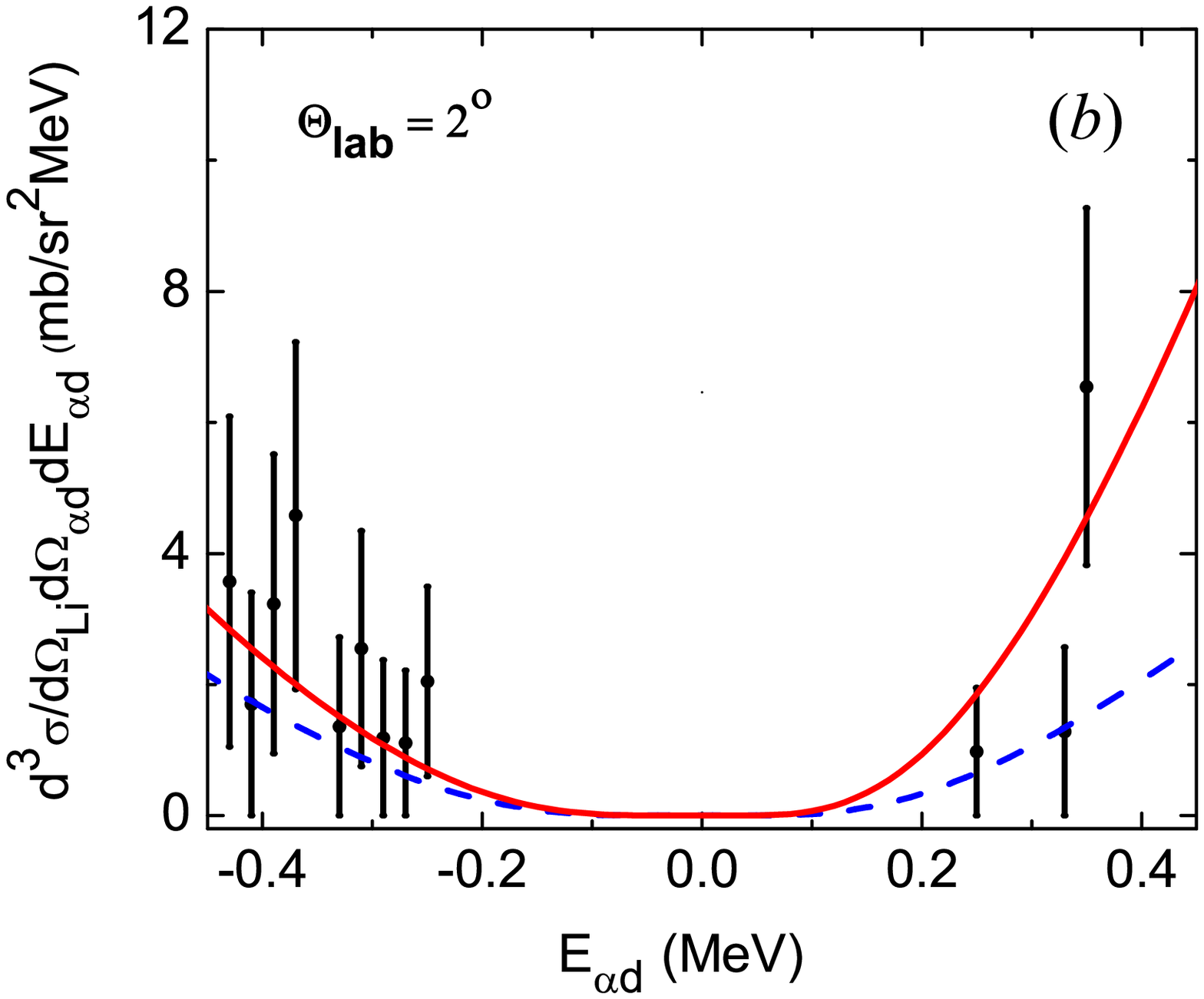}}
\parbox{6.7cm}{\includegraphics[width=6.7cm]{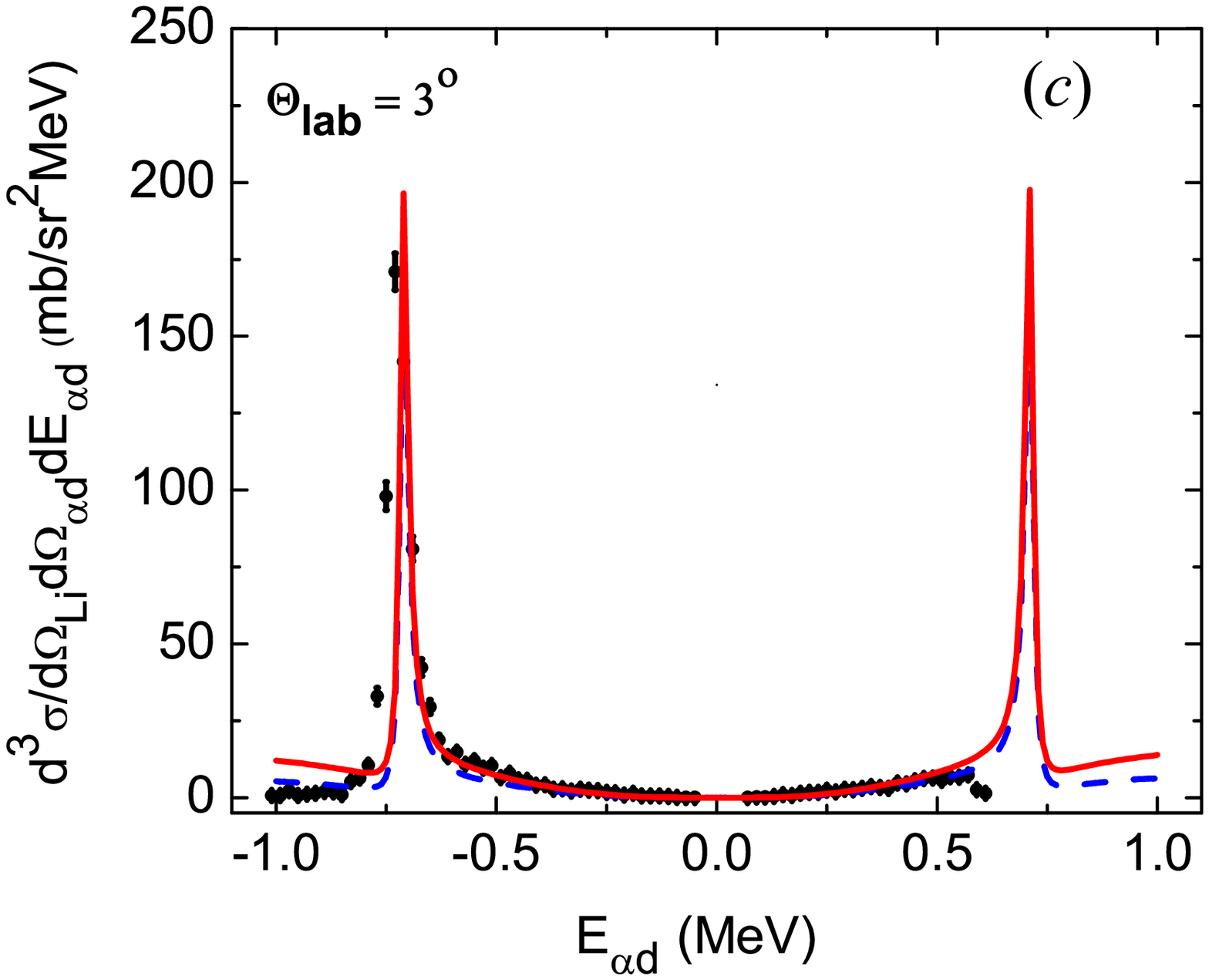}}
\parbox{6.7cm}{\includegraphics[width=6.7cm]{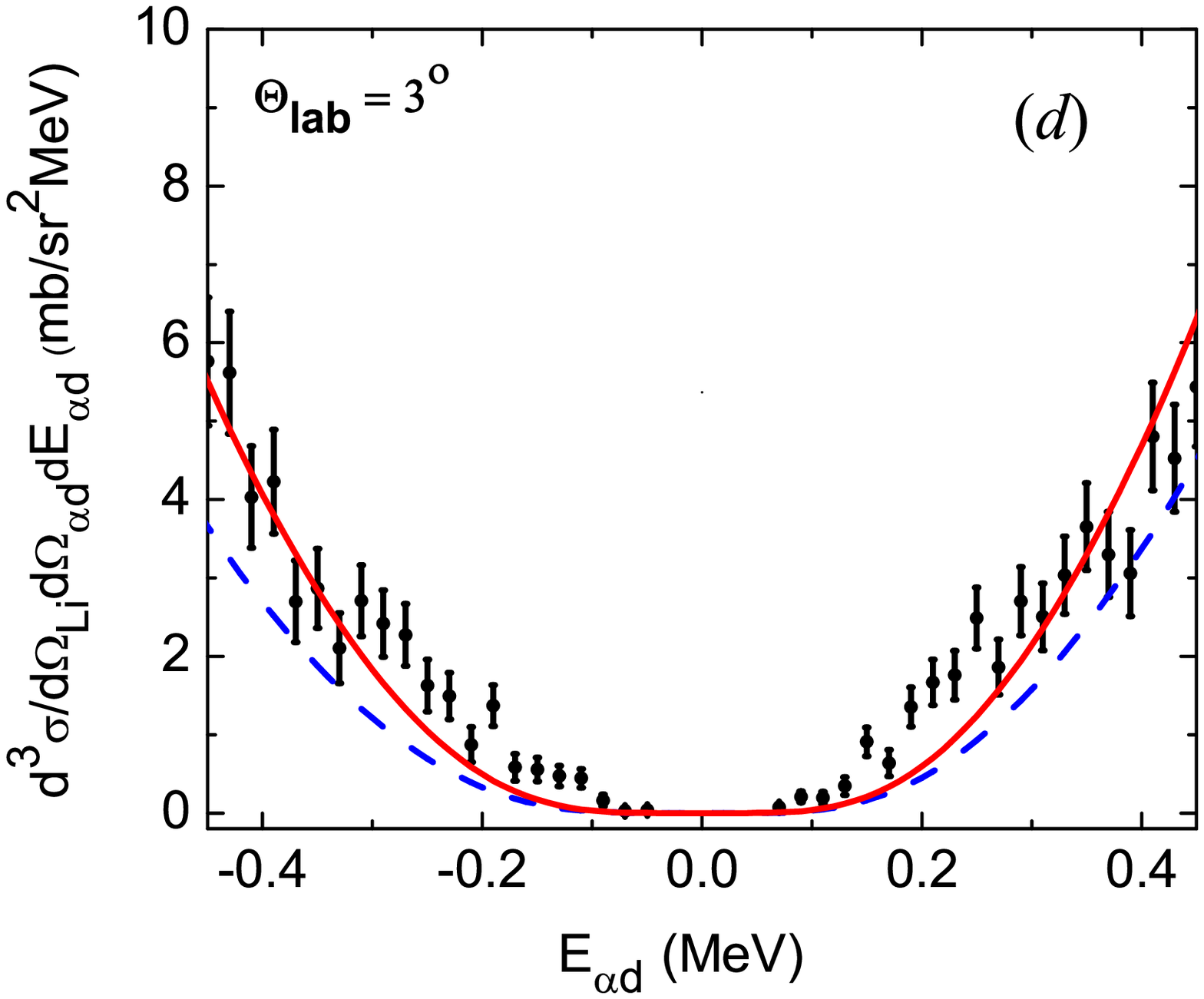}}
\end{center}
\caption{(Color online) Panel (\textit{a}): Energy distribution of
the total triple differential cross section for the $^6\rm{Li}$
breakup as a function of the relative energy $E_{\alpha d}$ at an
angle of $2^\circ$ (solid line). The distribution for the purely
Coulomb breakup is shown with dashed line. The experimental data
were taken from Ref. \cite{kin91}. Panel (\textit{b}): Same as in
(\textit{a}) but for energy in the close vicinity of zero.
Negative and positive $E_{\alpha d}$ energies denote backward and
forward emission, respectively, of the $\alpha$ particle in the
$^6\rm{Li}$ c.m. frame. Panels (\textit{c}) and (\textit{d}): Same
as in panels (\textit{a}) and (\textit{b}), respectively, but at
an scattering angle of $3^\circ$. \label{fig3}}
\end{figure}

\begin{figure}[thb]
\begin{center}
\parbox{6.7cm}{\includegraphics[width=6.7cm]{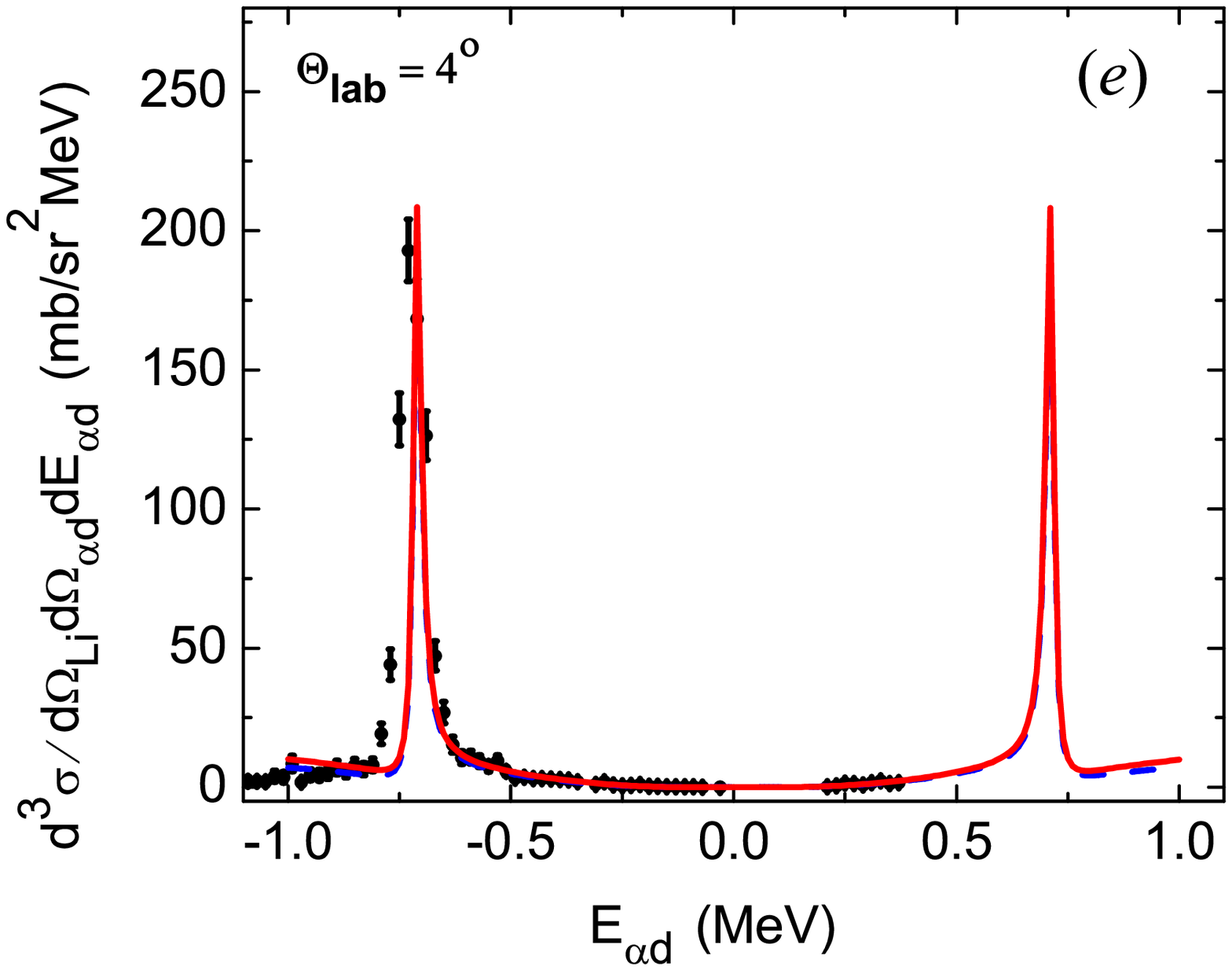}}
\parbox{6.7cm}{\includegraphics[width=6.7cm]{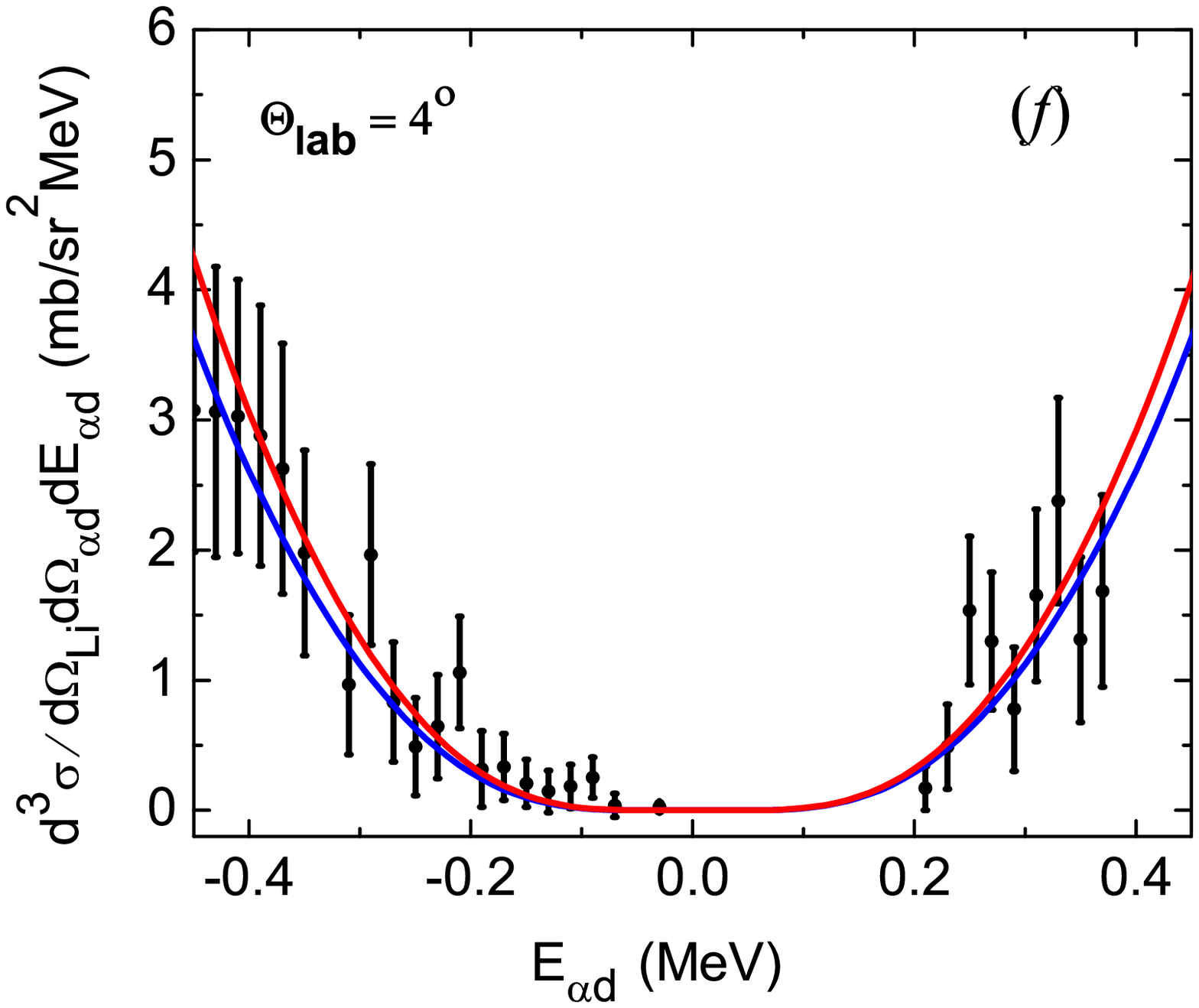}}
\parbox{6.7cm}{\includegraphics[width=6.7cm]{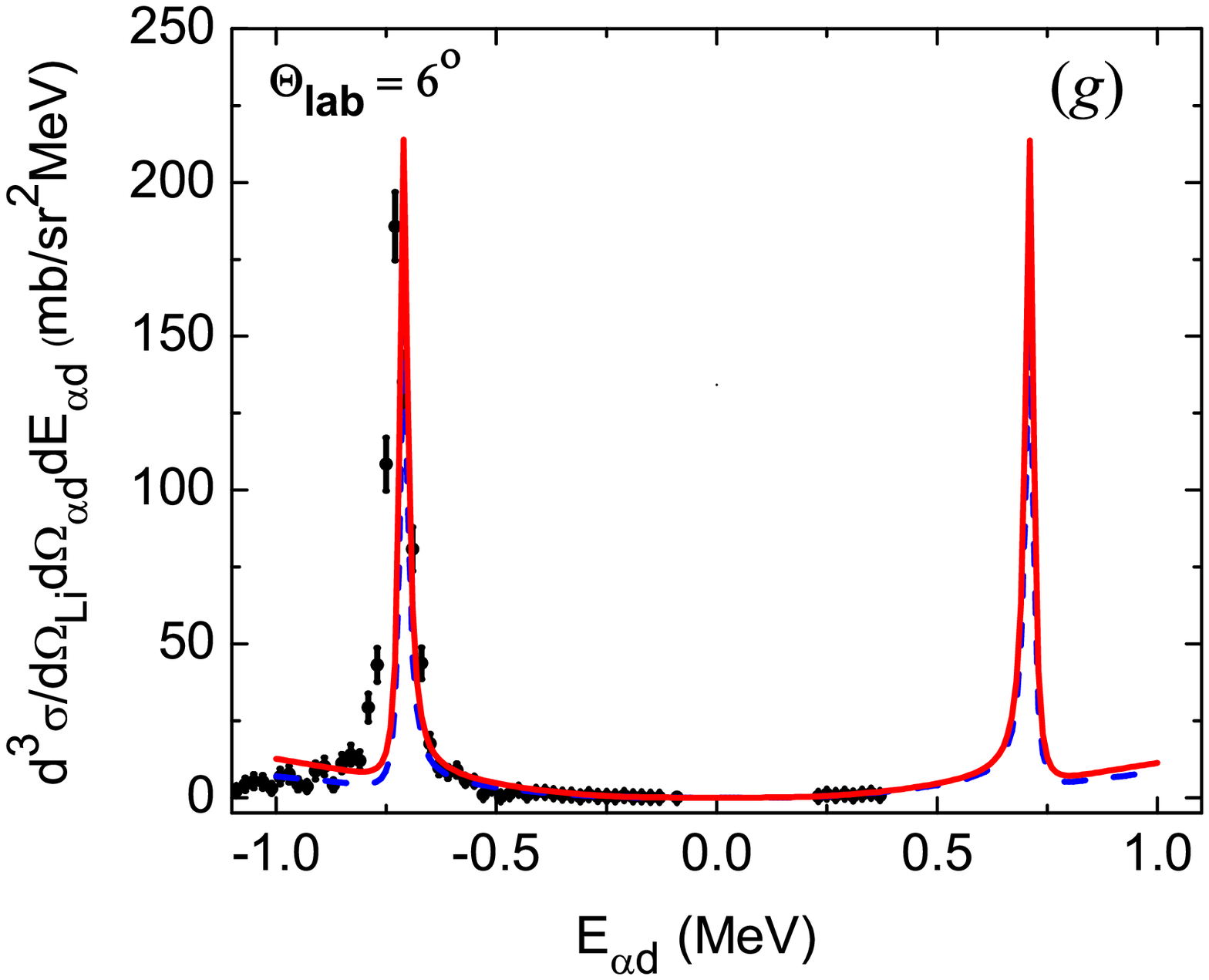}}
\parbox{6.7cm}{\includegraphics[width=6.7cm]{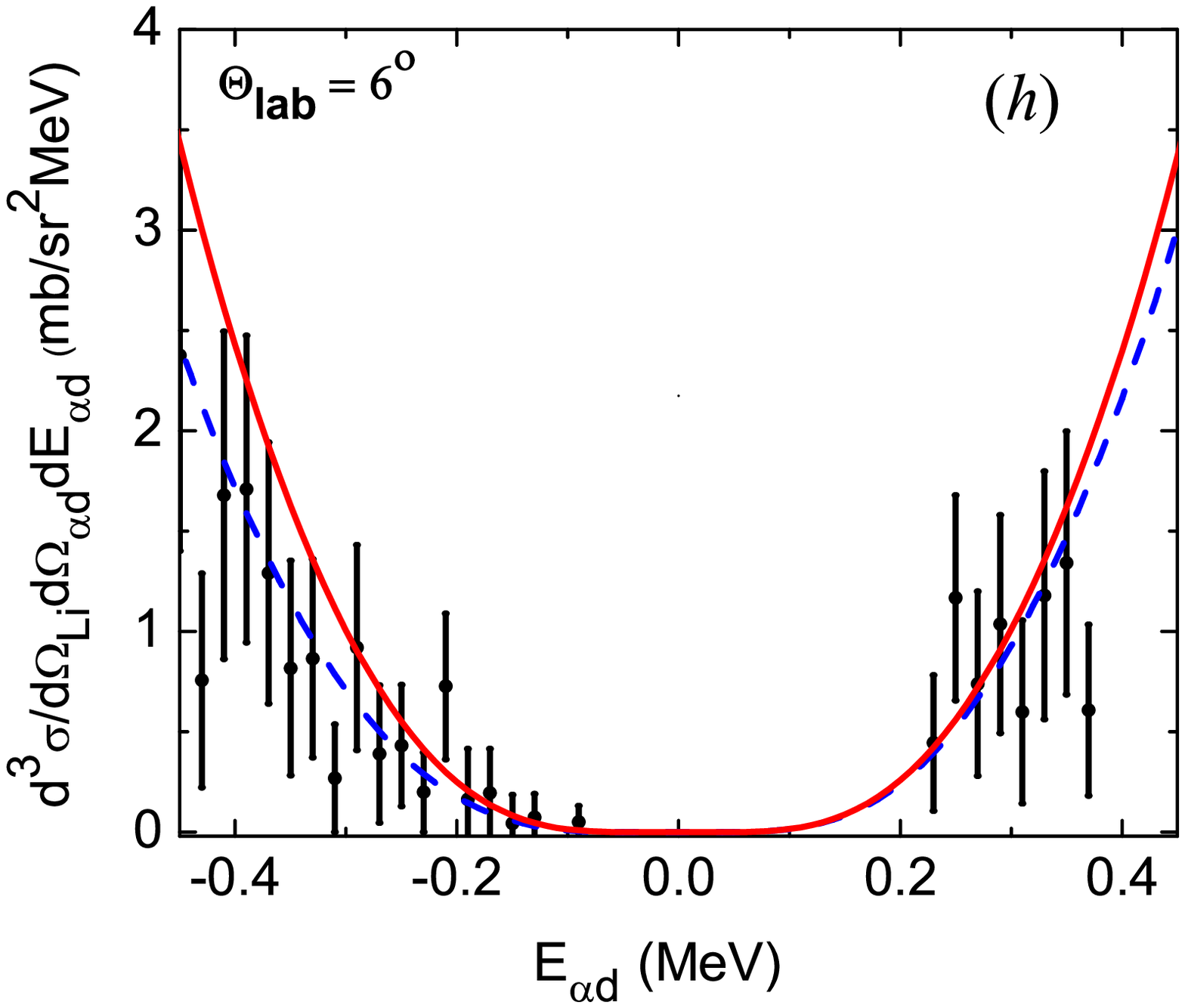}}
\end{center}
\caption{(Color online) Same as in Fig. \ref{fig3}. Panels
(\textit{e}) and (\textit{f}) show results at the scattering angle
$4^\circ$, while panels (\textit{g}) and (\textit{h}) show results
at the scattering angle $6^\circ$. \label{fig4}}
\end{figure}

\begin{figure}[thb]
\begin{center}
\parbox{8.6cm}{\includegraphics[width=8.6cm]{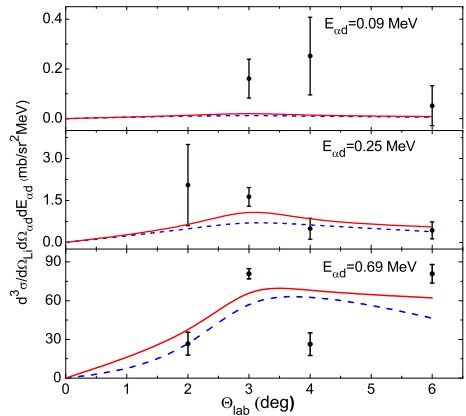}}
\end{center}
\caption{(Color online) Triple differential cross section for the
$^6\rm{Li}$ breakup as a function of the scattering angle
$\Theta_{\rm{lab}}$ at selected values of the relative $E_{\alpha
d}$ energy. Solid line is the sum of the Coulomb and nuclear
contributions, while dashed line is the purely Coulomb one. The
experimental data were taken from Ref. \cite{kin91}. \label{fig5}}
\end{figure}

The results of calculation of the triple cross sections are shown
in Figs.~\ref{fig3} and \ref{fig4}.
Comparison of our calculated triple cross sections  with the
experimental data shows that the Coulomb breakup gives main
contribution into disintegration process in the energy range 0.4
MeV $<E_{\alpha d}< 0.8$ MeV. A discrepancy exists with the
experimental data at the energies near zero and larger than 0.8
MeV for all scattering angles (particularly in the case
$\Theta_{\rm{lab}}=3^\circ$). As mentioned above the purely
Coulomb wave function for the $\alpha d$ continuum state maybe
applied when relative $\alpha d$ energy is close to zero. From
analyzing the results done by the authors of Refs. \cite{hess91,
kin91,maz03} at small values of the $E_{\alpha d}$ relative
energy, it is clear that the purely Coulomb breakup can not
explain the behavior of the cross section near the threshold.
However, the results of the authors of Ref. \cite{hamach10} showed
a huge contribution of the nuclear breakup at all $\alpha d$
energy region. Such a phenomenon maybe explained by the wide
angular region of the scattering angle of the c.m. of $^6\rm{Li}$
($0^\circ$-$5^\circ$) while the grazing angle at this experiment
is $\sim 2.5^\circ$-$3^\circ$. One expects the purely Rutherford
scattering only below $\sim 2^\circ$ (see also Figs. 7 and 10 of
Ref. \cite{hamach10}). At the grazing angle the projectile
$^6\rm{Li}$ moves along the trajectory tangential to the surface
of the target $^{208}\rm{Pb}$ where the nuclear breakup mostly
occurs. We note that an analysis of the Kiener \textit{et al.}
results concerning angular dependence $\Theta_{\rm{lab}}$ at very
low $E_{\alpha d}$ energy shows a large divergence with the
theoretical results following from the purely Coulomb
disintegration. The contribution of the nuclear breakup increases
the total cross section at most by a factor of 2 in the region
$E_{\alpha d}> 0.8$ MeV. A great disagreement still exists between
the theoretical and experimental results near zero $\alpha d$
relative energy. Figure \ref{fig5} demonstrates the dependence of
the total cross section and its Coulomb part on the scattering
angle at three values of $E_{\alpha d}$. Figure \ref{fig5} ($a$)
shows that the addition of the nuclear disintegration can not
decrease the disagreement with the experimental data at the
extremely low energies. At still higher energy the inclusion of
the nuclear breakup into theory might assist in achieving
reasonable agreement with the experimental data [Fig. \ref{fig5}
($b$) and Fig. \ref{fig5} ($c$)]. Accounting for the higher-order
effects (as mentioned in I) may lead to better agreement with the
experimental data at the extremely low energies. In the Kiener
\textit{et al.} work the grazing angle was $\sim 13^\circ$ while
the measurement was performed at the scattering angle range
$2^\circ$-$6^\circ$. Therefore the contribution of the nuclear
breakup to the differential cross section was small. The same
conclusion follows from the results of our calculations presented
in Table \ref{table1}. Note that similar behavior is observed for
other scattering angles.  When the energy $E_{\alpha d}$ becomes
larger than the resonant energy, the nuclear and Coulomb cross
sections become comparable to each other for the scattering angles
$2^\circ$ and $3^\circ$. However, at scattering angles $4^\circ$
and $6^\circ$, the Coulomb breakup cross section remains larger
than the nuclear breakup cross section.  Figure 3 of Ref.
\cite{hess91} shows how the Coulomb cross section approaches the
experimental data when the relative energy $E_{\alpha d}$
increases.

Figures \ref{fig3} and \ref{fig4} of our paper also depict the
existence of the backward-forward asymmetry for deuteron emission
in the $^6\rm{Li}$ c.m frame especially at smaller scattering
angles. Such a type of asymmetry was also observed in the Kiener
\textit{et al.} experiment \cite{kin91}. This asymmetry appears
due to the change of sign, both in the Coulomb and nuclear dipole
transition amplitudes, when the direction of deuteron emission is
changed from forward to backward in the $^6\rm{Li}$ c.m frame.

Thus, the results of our calculations confirms that the nuclear
contribution exists at the considered scattering angles and the
$\alpha d$ energies, but it is not as significant as that
concluded by the authors of Ref. \cite{hamach10}. Such a huge
contribution also contradicts the qualitative estimation given in
Ref. \cite{bert91}.
\begin{table}
\caption{Dependence of the ratio of the nuclear
($\sigma_{N}=\frac{d^3\sigma_N}{d\Omega_{\alpha
d}d\Omega_{\rm{Li}}dE_{\alpha d}}$) and the Coulomb
($\sigma_{C}=\frac{d^3\sigma_C}{d\Omega_{\alpha
d}d\Omega_{\rm{Li}}dE_{\alpha d}}$) triple cross sections for the
$^{208}\rm{Pb}(^6\rm{Li},\alpha\, d)^{208}\rm{Pb}$ breakup on
$E_{\alpha d}$ energy. The scattering angle $\Theta_{\rm{lab}}$ is
$3^\circ$. \label{table1}}
\begin{center}
\begin{tabular}{|c|c|c|c|} \hline
$E_{\alpha d}$  & $\sigma_{N}$  & $\sigma_{C}$  & $\sigma_{N}/ \sigma_{C}$ \\
(MeV) & $\rm{(mb/MeV/sr}^2)$ & $\rm{(mb/MeV/sr}^2)$ &  \\
\hline
1.00  & 6.774 & 6.316 & 1.072\\
0.95  & 6.372 & 5.921 & 1.076\\
0.90  & 5.932 & 5.376 & 1.104\\
0.85  & 5.459 & 4.641 & 1.176\\
0.80  & 4.981 & 3.809 & 1.308\\
0.77  & 4.806 & 3.968 & 1.211\\
0.76  & 4.847 & 4.708 & 1.030\\
0.75  & 5.048 & 6.632 & 0.761\\
0.74  & 5.663 & 11.73 & 0.483\\
0.73  & 7.555 & 26.98 & 0.280\\
0.72  & 13.97 & 79.58 & 0.175\\
0.71  & 24.61 & 172.8 & 0.142\\
0.70  & 16.56 & 111.9 & 0.148\\
0.65  & 5.043 & 18.10 & 0.279\\
0.60  & 3.757 & 10.49 & 0.358\\
0.55  & 3.007 & 7.621 & 0.394\\
0.50  & 2.395 & 5.854 & 0.409\\
0.45  & 1.856 & 4.503 & 0.412\\
0.40  & 1.378 & 3.372 & 0.409\\
0.35  & 0.964 & 2.400 & 0.402\\
0.30  & 0.621 & 1.581 & 0.393\\
0.25  & 0.354 & 0.927 & 0.382\\
0.20  & 0.168 & 0.453 & 0.371\\
0.15  & 0.058 & 0.162 & 0.360\\
0.10  & 1.07$\times 10^{-2}$& 3.060$\times 10^{-2}$& 0.349\\
0.05  & 3.05$\times 10^{-4}$& 9.047$\times 10^{-4}$& 0.337\\\hline
\end{tabular}
\end{center}
\end{table}

\section{Conclusion}

Using the simple two-body approach and the effective potential, we
have described the cross section and the astrophysical $S$ factor
of the $\alpha+d \to {}^6\rm{Li} +\gamma$ radiative capture. The
results are in good agreement with the known experimental data for
the range $ 0.4 < E_{\alpha d} < 3.0 $ MeV. For radiative capture
the contribution of the $E1$ transition to the cross section
becomes larger than the $E2$ one at an energy  less than 100 keV.
The calculated total value of the astrophysical $S$ factor equals
to $S(0)=1.22\,\,\rm{MeV}\,\rm{nb}$, while $S_{E1}(0)=
1.01\,\,\rm{MeV}\,\rm{nb}$ (83\% of the total $S$ factor) and
$S_{E2}(0)= 0.21\,\,\rm{MeV}\,\rm{nb}$ (17\% of the total $S$
factor). The results for the purely Coulomb breakup are in good
agreement with the known experimental data for the range $ 0.4 <
E_{\alpha d} < 0.8 $ MeV. This shows the validity of the
semiclassical method of the calculation for the Coulomb breakup at
high energy collision. The contribution of the $E1$ transition in
the Coulomb breakup is always less than the $E2$ one for all
energy regions. The nuclear disintegration is analyzed by
diffraction method which can be applied at small scattering
angles. The radius of ``black'' target $^{208}\rm{Pb}$ is taken as
a fit parameter and the application of the $\chi^2$ method gives
reasonable fit to the radius of the target. Our calculated nuclear
distortion is not large and suggests an overestimation of the
contribution of nuclear distortion made by Hammache \textit{et
al.} \cite{hamach10}. A comparison of our calculation of the
triple cross section, consisting of the Coulomb and nuclear parts,
with the experimental data, shows that disagreement still exists
for $E_{\alpha d}$ near zero. Taking into account the higher-order
effects may reduce this discrepancy at the low $\alpha d$ relative
energy ($E_{\alpha d}<0.3$ MeV). For instance, the three-body
Coulomb effects are known to strengthen with decreasing $\alpha d$
relative energy \cite{alt03,alt05}. Due to the existence of the
nuclear breakup and the small contribution of the $E1$ Coulomb
disintegration to the total cross section, it is impossible to
extract the correct value of the astrophysical $S$ factor for
$\alpha+d\to {}^6\rm{Li}+ \gamma$ radiative capture at low
$E_{\alpha d}$ energy. Extrapolation to zero energy can give the
value of the $E2$ component, if one is able to separate the
contributions of the nuclear disintegration  and the higher-order
effects from experimental data. To us, the simplest way to get an
accurate value of the astrophysical $S$ factor of peripheral
reactions, is to measure the ANC of the wave function in the bound
state with a high accuracy which governs the overall normalization
of the peripheral reaction cross section near zero energy.

\section*{Acknowledgments}
The authors are very grateful to A.\,M.~Mukhamedzhanov  for
fruitful discussions and useful comments.  The work was supported
by the HEC of Pakistan under Grant No. 20-1171 and 20-1283.

\end{document}